\documentclass[aps,prc,tightenlines,showpacs,superscriptaddress,twocolumn]
{revtex4-1}
\usepackage{graphicx}

\usepackage{soul}
\usepackage{color}
\usepackage[normalem]{ulem}

\begin{document}

\title{Deep Crustal Heating in a Multicomponent Accreted 
Neutron Star Crust}

\author{Andrew W. Steiner}
\affiliation{Institute for Nuclear Theory, University of Washington, 
Seattle, WA 98195}
\affiliation{Department of Physics \& Astronomy, National
Superconducting Cyclotron Laboratory, and the Joint Institute for
Nuclear Astrophysics, Michigan State University, East Lansing, MI}
\email{steiner3@uw.edu}

\begin{abstract}
A quasi-statistical equilibrium model is constructed to simulate the
multicomponent composition of the crust of an accreting neutron star.
The ashes of rp-process nucleosynthesis are driven by accretion
through a series of electron captures, neutron emissions, and
pycnonuclear fusions up to densities near the transition between the
neutron star crust and core. A liquid droplet model which includes
nuclear shell effects is used to provide nuclear masses far from
stability. Reaction pathways are determined consistently with the
nuclear mass model. The nuclear symmetry energy is an important
uncertainty in the masses of the exotic nuclei in the inner crust and
varying the symmetry energy changes the amount of deep crustal heating
by as much as a factor of two.
\end{abstract}


\preprint{INT-PUB-12-004}

\maketitle

\section{Introduction}

A large set of observational data from accreting neutron stars,
including X-ray bursts, superbursts, crust cooling, and the quiescent
luminosity of transiently accreting sources has become available in
recent years. This wealth of observational data leads to opportunities
to probe the properties of neutron stars, reactions on exotic nuclei,
and the nature of dense matter. However, the composition of the
deepest layers of the crust of accreting neutron stars is still not
yet well known. This work presents one of the first multicomponent
crust models which computes the properties of the deepest regions of
the crust.

In an accreting neutron star, as accreted matter (principally hydrogen
and helium) accumulates on the surface, nuclei in the outer crust are
pushed to deeper layers. At densities near $10^6$ g/cm$^{3}$, the
fusion of hydrogen and helium can become unstable and a thermonuclear
explosion results, an X-ray burst. These X-ray bursts generate heavier
nuclei by burning hydrogen, referred to as rp-process
nucleosynthesis~\cite{Schatz98}. As matter accretes, the burst ashes
are pushed to higher densities and undergo a series of nuclear
reactions: electron captures, neutron emissions, and pycnonuclear
fusions. These reactions drive the composition to nuclear statistical
equilibrium, the ground state in the neutron star crust. This stable
burning also generates heat of a few MeV per nucleon, which heats the
crust in addition to the unstable burning which occurs in X-ray
bursts. This is referred to as ``deep crustal
heating''~\cite{Haensel90}. It is this deep crustal heating which is
thought to drive the quiescent luminosity of accreting neutron
stars~\cite{Brown98}. At around $10^{11-12}$ g/cm$^{3}$ (the start of
the inner crust) the neutron separation energy becomes negative and
some of the neutrons form a quasi-free degenerate superfluid neutron
gas. The neutron emissions dominate the deep crustal heating at
densities just above the density at which the inner crust begins.
Finally, near $10^{14}$ g/cm$^{3}$, the crust ends when nuclei are no
longer energetically favorable.

One success in connecting the observations to theoretical models is
the work in Refs.~\cite{Shternin07,Brown09}, which used a theoretical
cooling model to describe the crust cooling of KS 1731 and MXB 1659.
The X-ray flux from these objects was observed immediately after
outburst, and this flux decreased according to a broken power-law: the
flux decreases more weakly with time at early times before the photons
from the inner crust have reached the photosphere, and the flux
decreases more strongly with time at later times. This effect is due
principally to the larger thermal conductivity from superfluid
quasi-free neutrons in the inner crust (it is the first definitive
observation of superfluidity in the crust). Ref.~\cite{Brown09} 
showed that the crust must have a relatively small impurity
parameter to have a thermal conductivity large enough to reproduce the
data.

Another success is the observation of the crustal cooling of SAX J1808
after an accretion outburst. SAX J1808 is a neutron star with
transiently accretes from a small main sequence companion. These
accretion events warm up the neutron star crust relative to the core
and the cooling of the crust can be observed right after the end of an
accretion event. Ref.~\cite{Heinke06} compared the observations to
models developed in Ref.~\cite{Yakovlev04} which required the
composition and heating of the accreted crust as an input. The result
was that the cooling of SAX J1808 was so rapid as to suggest that the
minimal neutron star cooling model~\cite{Page04,Page09} was
insufficient to explain the strong decrease in luminosity. This
suggests that extra cooling beyond the minimal model, such as direct
Urca or the cooling from pions or quarks, is present in SAX J1808.

Some observations have been difficult to reproduce with current crust
models. An instability to Carbon fusion at around $10^{10}$ g/cm$^{3}$
may generate a superburst~\cite{Cumming01,Strohmayer02}, an energetic
form of X-ray burst which is observed in some sources. This
instability is strongly temperature dependent. Current models suggest
that the crust is too cold do destabilize Carbon fusion, thus
suggesting superburst models require revisiting
\cite{Brown04,Cumming06,Gupta07,Brown09}. While the carbon fusion cross
section is not well understood at these low energies, a severe
enhancement in the fusion rate would be required to explain many
superbursts~\cite{Cooper09}. One important input parameter in these
models is the amount of heating in the inner crust. If the heating in
the inner crust was sufficiently strong, this could alleviate
difficulties in current superburst models.

Theoretical models of X-ray bursts which couple reaction networks to
hydrodynamics and radiation transport have met with considerable
success~\cite{Woosley04}. These complicated computational frameworks
cannot yet follow the reactions all the way the higher densities
probed in superbursts, in part because of the computational cost of
the immense nuclear reaction networks which are required. An
alternative was employed in Refs.~\cite{Gupta07,Gupta08}, which uses a
full reaction network with simplified hydrodynamics. These authors
found extra heating from electron captures into excited states which
partially, but not fully, alleviated the problem with superburst
models. These simplified models become difficult in the inner crust,
where nuclear reactions in the medium of quasi-free neutrons are not
well known. Thus, these models cannot yet trace the evolution of
accreted material as it sinks into the inner crust. A final difficulty
is the requirement that the nuclear reactions are consistent with the
nuclear masses, themselves subject to significant uncertainties at the
relevant densities.

On the other hand, multicomponent models of the accreted crust are
particularly important because crusts are at a sufficiently low
temperature and density to strongly limit the possible nuclear
reactions which might occur. Nuclear structure effects also come into
play, and some reactions which would not have been possible in a
one-component system open up new reaction flows in a multicomponent
system.

A simplification comes from the fact that, in the deepest regions of
the crust, most of the relevant nuclear reaction rates are strongly
density dependent. This means that as soon as a reaction channel
becomes energetically allowed, its rate rises and quickly becomes much
faster than the local accretion timescale. A quasi-statistical
equilibrium (QSE) ensues where, to a good approximation, electron
captures and neutron emissions always proceed so long as they are
exothermic. This QSE state is almost independent of the details of the
nuclear reactions and depends more strongly on the masses of the
nuclei which are present at any particular density (i.e. the Q
values). Pycnonuclear fusion reactions are allowed when their fusion
timescale is much faster than the local accretion timescale and can
be handled separately.

In this work, a one-zone multicomponent QSE model is used to describe
the composition and heating in the crust of an accreting neutron star.
When a particular nuclear reaction is energetically disfavored, its
rate is presumed to be zero. When non-fusion reactions are
energetically favored, their rate is taken to be infinite, proceeding
until they become energetically disfavored again. Pycnonuclear fusion
is handled by taking advantage of recent work on the relevant nuclear
S-factors. Nuclear masses are described with a modern liquid droplet
model which contains corrections for nuclear shell effects and matches
available experimental data with an accuracy near that of more
microscopic approaches. In-medium corrections are also included, i.e.
the masses of each nuclei depend on the temperature, the potential
presence of quasi-free neutrons, and also on the ambient electron
density. The use of a liquid droplet model allows the inclusion of
almost all of the relevant physics, but avoids the large computational
time of more microscopic models.

\section{Crust model}

This work assumes that the multicomponent crust is uniformly mixed,
that nuclei are randomly distributed and uncorrelated, except for the
lattice correlations which naturally occur in the Coulomb solid. The
Wigner-Seitz approximation is used, and each nucleus occupies one and
only one Wigner-Seitz cell, filled with electrons and quasi-free
neutrons. Each cell is fixed in size by requiring it have no overall
electric charge. 

The assumption of uniform mixing may be a poor one, especially at
lower densities. Recent molecular dynamics
calculations~\cite{Horowitz07,Horowitz08b,Horowitz09} suggest that
lighter and heavier nuclei separate. This will not affect the bulk
energetics of the system but may have a significant impact on the
pycnonuclear reaction rates described below. However, these
microscopic simulations are computationally difficult, and cannot yet
be performed for a wide range of density regimes, nuclear mass models,
and initial compositions. These simulations also employ very
simplified models of the nucleon-nucleon interaction and do not handle
all of the potential finite-size effects. 

\subsection{In-medium nuclear mass formula}

The binding energy of the nucleus in medium with index $i$ can be
written as a sum of terms,
\begin{eqnarray}
E_{\mathrm{nuc}}
(N_i,Z_i,n_{n,\mathrm{out}},n_{p,\mathrm{out}},n_e,T) = E_{\mathrm{bulk}} +
E_{\mathrm{surf}} + \nonumber \\
E_{\mathrm{Coul}} + E_{\mathrm{pair}} + E_{\mathrm{shell}} \, .
\end{eqnarray}
For simplicity, $\hbar = k_B = c = 1$ in the following. 
The quantities $n_{n,\mathrm{out}}$ and $n_{p,\mathrm{out}}$ 
denote the average quasi-free neutron and proton densities outside
of nuclei. The demarcation of nucleons inside and outside of nuclei
is clearly artificial, but has proven to be a good approximation
for the bulk thermodynamic properties of matter except at the highest
densities in the crust. The dependence of the mass formula on the
$n_{n,\mathrm{out}}$ and $n_{p,\mathrm{out}}$, and on the ambient
electron density, $n_e$, is required to reflect the fact that the
masses of nuclei depend on the surrounding medium. While the physical
properties (radii, internal neutron and proton densities, etc.) of
each nucleus in the distribution are different for each nucleus and
should thus be given a subscript $i$, this subscript is omitted in
this section to simplify the notation. The eleven nuclear mass model
parameters described below are all identical for each nucleus, and the
distinction between model parameters and nuclear properties will be
made clear.

The bulk part of the nuclear energy is constructed from a given
equation of state (EOS) of bulk nuclear matter, denoted
$\varepsilon_{\infty}(n_n,n_p,T)$. Let 
${\tilde \varepsilon}_{\infty}(n_n,n_p,T)$ denote
the energy density of bulk matter defined without the rest mass
energy density, and similarly for the Helmholtz free energy density,
${\tilde f}$, e.g.
\begin{eqnarray}
{\tilde \varepsilon}_{\infty}(n_n,n_p,T) &=& 
\varepsilon_{\infty}(n_n,n_p,T) - m_n n_n - m_p n_p 
\nonumber \\
{\tilde f}_{\infty}(n_n,n_p,T) &=& 
f_{\infty}(n_n,n_p,T) - m_n n_n - m_p n_p
 \nonumber \\
& =&  {\tilde \varepsilon}_{\infty}(n_n,n_p,T) - T
s(n_n,n_p,T) 
\end{eqnarray}
The function $\varepsilon_{\infty}(n_n,n_p,T)$ is given either by a
Skyrme~\cite{Skyrme59} model or by the model of Akmal et
al~\cite{Akmal98} (hereafter APR). Skyrme models
SLy4~\cite{Chabanat95}, Gs, and Rs~\cite{Friedrich86} are used,
motivated by the fact that they provide a variation in the symmetry
energy while still giving reasonable saturation properties and neutron
star masses and radii. Finite temperature corrections in the bulk part
of the nuclear energy are negligible at the temperatures of interest
($<10^9$ K), so $s(n_n,n_p,T)$ will be taken to be zero. In effect,
the mass model described below is actually an entire class of mass
models with different functions, $\varepsilon_{\infty}$, all of which
have a comparable quality as evaluated by the RMS deviation of the
mass excess, yet with different compressibilities and symmetry
energies. This allows one to estimate the uncertainties in the
property of the crust due to the uncertainties in the nature of the
nucleon-nucleon interaction~\cite{Oyamatsu07,Steiner08,Newton11}.

An approximate expression for the volume of the nucleus is
$V_{\mathrm{nuc}}=A/n_{B,\mathrm{in}}$, where $n_{B,\mathrm{in}}
\equiv n_{n,\mathrm{in}} + n_{p,\mathrm{in}}$, the sum of the average
neutron and proton densities inside the nucleus. Then the bulk energy
is (c.f.~\cite{Steiner08}),
\begin{equation}
E_{\mathrm{bulk}} = {\tilde \varepsilon}_{\infty} (n_{n,\mathrm{in}},
n_{p,\mathrm{in}},T) \left(\frac{A}{n_{B,\mathrm{in}}}\right) \, ,
\label{eq:bulk}
\end{equation}
where $n_{n,\mathrm{in}}$ and $n_{p,\mathrm{in}}$ are the internal
neutron and proton densities inside the nucleus. In the present model,
the internal baryon density is chosen to be
\begin{equation}
n_{B,\mathrm{in}} = n_0 + n_1 I^2 \, ,
\label{eq:n}
\end{equation}
where $n_0$ and $n_1$ are parameters of the model and $I \equiv
1-2Z/A$. The individual average neutron and proton number densities
are given by
\begin{eqnarray}
n_{n,\mathrm{in}} & = & n_{B,\mathrm{in}} (1 + \delta) / 2 + g(\chi)
\nonumber \\
n_{p,\mathrm{in}} & = & n_{B,\mathrm{in}} (1 - \delta) / 2 - g(\chi)
\label{eq:np}
\end{eqnarray}
and the density asymmetry $\delta= \zeta I $ where $\zeta$ is an
additional parameter of the model. Other models choose to treat
$n_{n,\mathrm{in}}$ and $n_{p,\mathrm{in}}$ as nucleus shape
parameters to be minimized over for each nucleus, but this does not
typically improve the fit to experimentally measured masses. If
$\zeta=1$, nuclei in vacuum have no neutron skin and $\zeta<1$
indicates the presence of a neutron skin. The fraction of the
Wigner-Seitz cell volume occupied by the nucleus, $\chi$, is described
below. Without the additional correction, $g(\chi)$, the energy of
ultra-neutron rich matter the densities implied by $n_{n,\mathrm{in}}$
and $n_{p,\mathrm{in}}$ fail to give a physical value of the
core-crust transition density as computed in, e.g.
Ref.~\cite{Oyamatsu07}. The function $g(\chi)$ is defined by
\begin{equation}
g(\chi) \equiv f_{nC} \left(\frac{1-e^{f_{nE} \chi}}{1 -
  e^{f_{nE}}}\right) \, ,
\end{equation}
where $f_{nE} \equiv 5$ and $f_{nC} \equiv 1/2$ alleviates this
difficulty. This choice of this functional form for $g(\chi)$ is
purely phenomenological; the exponential in $\chi$ ensures that nuclei
are only affected at the deepest regions of the crust. Nuclear radii
are defined by the relations $4 \pi R_n^3 n_{n,\mathrm{in}} = 3N$ and
$4 \pi R_p^3 n_{p,\mathrm{in}} = 3Z$.

The radius of the Wigner-Seitz cell, $R_{\mathrm{WS}}$ for each
nucleus is determined by assuming that each cell contains the same
number of protons and electrons, i.e.
\begin{equation}
\frac{4}{3} \pi R_{\mathrm{WS}}^3 n_e = Z
\end{equation}
where the electron density $n_e$ is taken to be the same in the WS
cells of all nuclear species. This choice approximately ensures that
the edges of every Wigner-Seitz cell is at a fixed electrostatic
potential. Furthermore, this choice ensures that the WS cells for each
nucleus in the multicomponent mixture occupy the entire volume and
that each neutron in the quasi-free neutron gas is associated with one
and only one WS cell. The volume fraction of the cell which is
occupied by the neutrons in the nucleus is $\chi \equiv
(R_n/R_{\mathrm{WS}})^3$ and the volume fraction occupied by protons
is $\chi_p \equiv (R_p/R_{\mathrm{WS}})^3$. Because the size of the
cell depends on the ambient electron density, and because the Coulomb
energy in each nucleus depends on the size of the cell, {\it the mass
  of every nucleus depends (albeit weakly) on the number density of
  every other species in the system}. This is expected, since the
Coulomb energy has longer range than the nuclear forces and couples
each nucleus to the others. Defining volume of each cell, $V_i = 4 \pi
R_{\mathrm{WS},i}^3/3$ ensures that the identity $\sum_i n_i V_i = 1$ 
exactly holds. 

The surface energy is given as
\begin{equation}
E_{\mathrm{surf}} = 4 \pi R_{\mathrm{surf}}^2 \sigma {\cal B}(n_n,n_p)
\end{equation}
where $R_{\mathrm{surf}}$ is defined by the relation
\begin{equation}
\frac{4}{3} \pi R_{\mathrm{surf}}^3 n_{B,\mathrm{in}} = A \, ,
\end{equation}
the quantity ${\cal B}$ is defined by
\begin{equation}
{\cal B}(n_n,n_p) \equiv \frac{16 + b}{\left[1/x^3 + b + 1/(1-x)^3
    \right]} \, ,
\end{equation}
and $\sigma_{\delta}=96/(b+16)$. This is equivalent to Eq.~5 in
Ref.~\cite{Steiner08}, and ensures that the surface energy properly
obeys the $x^3$ dependence shown in Ref.~\cite{Lattimer85} to match
Thomas-Fermi calculations of very neutron-rich nuclei. It has the
consequence that nuclei at large densities have small surface energies
because the vanishing proton fraction requires the neutron density
distribution to be quite diffuse (even though the current model
contains no explicit diffusiveness).

The Coulomb energy is
\begin{equation}
E_{\mathrm{Coul}} = 2 {\cal C} \pi e^2 R_p^2 
\left(n_{p,\mathrm{in}}-n_{p,\mathrm{out}}\right)^2 f_d(\chi_p) 
\left(\frac{A}{n_{B,\mathrm{in}}}\right)
\end{equation}
where the function $f_c(\chi_p)$ is given by
\begin{equation}
f_d (\chi_p) = \frac{1}{d+2} \left\{\left(\frac{2}{d-2}\right)
\left[1 - \frac{1}{2}
\chi_p^{\left(1-2/d\right)} \right] + \chi_p \right\}
\end{equation}
where $d$ is the dimensionality (shape) of the nucleus. All nuclear
are assumed to be spherical, i.e. $d=3$. The coefficient, ${\cal C}$, is
an arbitrary parameter which decreases the Coulomb energy slightly, in
order to model the diffusiveness of the proton distribution in
laboratory nuclei. This formula can include pasta by allowing $d$ to
be different from 3, but this possibility is left to future work.
This differs slightly from the original expression
in~\cite{Ravenhall83}: the factor
$\left(n_{p,\mathrm{in}}-n_{p,\mathrm{out}}\right)^2$ ensures that the
Coulomb energy vanishes if the proton densities internal and external
to nuclei are equal. In practice, however, this correction will not
affect our results.

The pairing energy is
\begin{equation}
E_{\mathrm{pair}} = \left\{
\begin{array}{ll}
- A^{-1/3} E_{\Delta} & \mathrm{N~and~Z~even} \\
A^{-1/3} E_{\Delta} & \mathrm{N~and~Z~odd} \\
0 & \mathrm{otherwise} 
\end{array}
\right. \, .
\end{equation}
The exponent $1/3$ is known to be not well constrained by fitting to
laboratory nuclei, and varying this exponent as a model parameter does
not substantially improve the fit.

For the shell energy, the corrections described in
Ref.~\cite{Dieperink09} are used, modified to correct for the medium.
Shell corrections can be particularly difficult to evaluate in quantum
mechanical models of nuclei at these densities because of spurious
shell effects which are generated by the boundary of the Wigner-Seitz
cell~\cite{Grasso08}. It is also unclear how to properly
modify these shell effects for the medium. One expects that, as the
number density of neutrons outside of nuclei increases relative to the
number density of neutrons inside nuclei, the shell effects become
less pronounced. This is treated phenomenologically by applying 
quenching functions
\begin{eqnarray}
\Lambda_n &=& \left( \frac{n_{n,\mathrm{in}}-n_{n,\mathrm{out}}}
{n_{n,\mathrm{in}}} \right)^2
\nonumber \\
\Lambda_p &=& \left( \frac{n_{p,\mathrm{in}}-n_{p,\mathrm{out}}}
{n_{p,\mathrm{in}}} \right)^2 
\end{eqnarray}
to the shell correction energy, i.e. 
\begin{eqnarray}
S_2 &=& \frac{n_v {\bar n}_v}{D_n} \Lambda_n + 
\frac{z_v {\bar z}_v}{D_z} \Lambda_p \nonumber \\
S_3 &=& \frac{n_v {\bar n}_v \left(n_v - {\bar n}_v\right)}{D_n}
\Lambda_n + \frac{z_v {\bar z}_v \left(z_v - 
{\bar z}_v\right)}{D_z} \Lambda_p \nonumber \\
S_{\mathrm{np}} &=& \frac{n_v {\bar n}_v z_v {\bar z}_v}{D_n D_z} 
\Lambda_n \Lambda_p
\end{eqnarray}
Then the final shell correction is 
\begin{equation}
E_{\mathrm{shell}} = a_1 S_2 + a_2 S_2^2 + a_3 S_3 + a_{\mathrm{np}}
S_{\mathrm{np}}
\end{equation}
The neutron magic numbers are set to 2, 8, 14, 28, 50, 82, 126, 184,
228, 308, and 406, as suggested by Ref.~\cite{Denisov04}, and the
proton magic numbers are all within the range accessible by
experiment.

In summary, the 11 free parameters in this model (outside of the
input equation of state of bulk nuclear matter which is a kind of
parameter in itself) are the surface tension in MeV/fm$^{2}$,
$\sigma$, the surface symmetry energy $\sigma_{\delta}$, the
correction factor to the Coulomb energy, ${\cal C}$, the asymmetry
parameter $\zeta$, the central density parameters, $n_0$ and $n_1$
which are expressed in units of fm$^{-3}$, the pairing energy
$E_{\Delta}$, and the four parameters for the shell effects, $a_1$,
$a_2$, $a_3$, $a_{\mathrm{np}}$.

\begin{table}
\begin{tabular}{ccccc}
Quantity (Units) & APR & SLy4 & Gs & Rs \\
\hline
$n_0$ (fm$^{-3}$) & 0.1786 & 0.1789 & 0.1479 & 0.1504 \\
$n_1$ (fm$^{-3}$) & -0.1057 & -0.08760 & 0.03355 & 0.02920 \\
$\eta$ & 0.8804 & 0.8798 & 0.8642 & 0.8696 \\
$\sigma$ (MeV/fm$^{2}$) & 1.155 & 1.154 & 0.9772 & 0.9906 \\
$\sigma_{\delta}$ & 1.382 & 1.251 & 0.4446 & 0.4597 \\
${\cal C}$ & 0.8957 & 0.8933 & 0.9246 & 0.9227 \\
$E_{\Delta}$ (MeV) & 5.224 & 5.226 & 5.213 & 5.218 \\
$a_1$ (MeV) & -1.390 & -1.390 & -1.378 & -1.373 \\
$a_2$ (MeV) & 0.008931 & 0.01001 & 0.01300 & 0.01264 \\
$10^3$ $a_3$ (MeV) & 2.380 & 2.360 & 1.865 & 1.920 \\
$a_{\mathrm{np}}$ (MeV) & 0.1133 & 0.1137 & 0.09897 & 0.09944 \\
\hline
$\delta m_{\mathrm{RMS}}$ (MeV) & 1.132 & 1.124 & 1.228 & 1.200 \\
\end{tabular}
\caption{The model parameters and RMS deviation in the mass excess for
  the homogeneous equations of state used in this work.
}
\label{tab:massparms}
\end{table}

The results of the fit to the experimental mass data from
Ref.~\cite{Audi03} are given in Table~\ref{tab:massparms}. Note that,
because of the inclusion of shell effects, the quality of the mass
formula is about 1.2 MeV, much closer to the 0.7 MeV deviation
observed for FRDM, and much improved from a typical liquid droplet
model which has a deviation of 2.6 MeV or more. It is also instructive
to see how the parameters depend with the density dependence of the
symmetry energy: APR and SLy4 have symmetry energies which depend
rather more weakly with density and Gs and Rs have symmetry energies
which depend more strongly with density. It is clear that the surface
symmetry energy $\sigma_{\delta}$ is correlated with the symmetry
energy, as is the parameter $n_1$, but the pairing and shell
parameters are only weakly correlated to the symmetry energy.  

\subsection{Free energy of multi-component matter}

The free energy density of matter (without the rest mass energy
density) is
\begin{eqnarray}
&{\tilde f}\left(\{n_i\},n_{n,\mathrm{out}},n_{p,\mathrm{out}},T\right) = 
& \nonumber \\
&\sum_i \left[E_{\mathrm{nuc}}
(Z_i,N_i,n_{n,\mathrm{out}},n_{p,\mathrm{out}},n_e,T) 
n_i + f_C(n_i,T) \right] &\nonumber \\
&+ (1 - \phi) {\tilde f}_{\infty}(n_{n,\mathrm{out}},n_{p,\mathrm{out}},T) + 
{\tilde f}_{\mathrm{elec}}(n_e,T)&
\label{eq:totfr}
\end{eqnarray}
where $f_C$ is the classical expression for the free energy density of
nucleus $i$ with density $n_i$ at temperature $T$ and
$f_{\mathrm{elec}}$ is the free energy density of electrons without
the electron rest mass energy density. One can also in principle
include finite temperature corrections to $f_C$ appropriate to dense
matter as described in Ref.~\cite{Souza09}. The contribution from
$f_C$ is small and will be omitted here. Other thermodynamic
quantities, such as the Gibbs energy density (see the Appendix), can
be trivially obtained from the free energy density in the usual way.
Hereafter, the proton drip is assumed to be negligible, i.e.
$n_{p,\mathrm{out}}=0$. The electron density is not independent, $n_e$
must be self-consistently determined from the nuclear densities, i.e.
$n_e = \sum_i n_i Z_i$. The partial volume fraction
\begin{equation}
\phi\equiv
\sum_i \phi_i \equiv \sum_i \frac{4}{3} \pi R_{n,i}^3 n_i
\end{equation}
is also not independent and is a function of $\{n_i\}$. In this
formulation, $n_{n,\mathrm{out}}$ is the local number density of the
quasi-free neutron gas in between nuclei, while
$n_{n,\mathrm{out}}(1-\phi)$ is the number of neutron per unit volume
inside a large volume of many Wigner-Seitz cells. In a one-component
system, $\phi = \chi$. 

The rest mass part of the energy density (not included above) is
\begin{eqnarray}
\rho \equiv \varepsilon_{\mathrm{rest}} &=& 
\sum_i \left( N_i m_n n_i + Z_i m_p n_i \right) + 
\nonumber \\
&& \left( 1 - \phi \right) n_{n,\mathrm{out}} m_n + m_e n_e \, .
\label{eq:rest}
\end{eqnarray}
Note that the rest mass energy density is defined in terms
of neutron and proton degrees of freedom, even though the protons
are typically bound in nuclei, which is convenient since 
nuclear binding energies are being modified by the medium.

The excluded volume correction can be written
\begin{equation}
f_{\mathrm{exc}} = - \sum_i \phi_i
{\tilde f}_{\infty}(n_{n,\mathrm{out}},0,T) 
\end{equation}
One can think about this either as a correction to the total energy
density as is written in Eq.~\ref{eq:totfr} or as a correction to the
bulk energy of each nucleus
\begin{equation}
E_{\mathrm{exc},i} = - \left(\frac{\phi_i}{n_i}\right) {\tilde f}_{\infty}
(n_{n,\mathrm{out}},0,T) \, .
\end{equation}
When written this way, one is effectively defining the mass of the
nucleus in the medium relative to energy of pure neutron matter in an
WS cell of equivalent volume, rather than defining the nuclear mass
relative to the vacuum. (This is similar to what has been historically
done in some Hartree-Fock and Thomas-Fermi calculations in order 
to remove spurious
shell effects~\cite{Bonche84,Suraud87}) If one defines
\begin{equation}
N^{\prime}_i = N_i - \phi_i \frac{n_{n,\mathrm{out}}}{n_i} = 
N_i - \frac{4}{3} \pi R_{n,i}^3 n_{n,\mathrm{out}}
\end{equation}
and define ${\cal E}_{\mathrm{nuc},i} = E_{\mathrm{nuc},i} +
E_{\mathrm{exc},i}$ then the free energy density of matter
can be rewritten
\begin{eqnarray}
&f\left(\{n_i\},n_{n,\mathrm{out}},0,T\right) = 
& \nonumber \\
&\sum_i n_i \left({\cal E}_{\mathrm{nuc},i} + 
Z_i m_p + N^{\prime}_i m_n \right) & \nonumber \\
&+ f_{\infty}(n_{n,\mathrm{out}},0,T) + f_{\mathrm{elec}}(n_e,T)& \, .
\label{eq:newfr}
\end{eqnarray}
This form now explicitly includes the rest mass part of the energy
density and also makes computing analytical derivatives of the free
energy a bit simpler. The factor of $(1-\phi)$ from the neutron free
energy in Eq.~\ref{eq:totfr} is no longer present, because the
excluded volume correction has been absorbed into the definition of
${\cal E}_{\mathrm{nuc},i}$. Note that some authors also refer to
finite-volume corrections to $f_C(n_i,T)$ as excluded volume
corrections, but these corrections are negligible here. The total
baryon density is
\begin{equation}
n_B = \sum_i A_i n_i + (1 - \phi) n_{n,\mathrm{out}}
\end{equation}

This formulation (with some minor additional finite-temperature
effects) of the properties of dense matter is rich enough to express
the properties of stellar matter in thermodynamic equilibrium at
higher densities, conditions relevant for Type II supernovae. In this
case, the multicomponent nature has a small impact on the overall
thermodynamic properties~\cite{Lattimer85} thus justifying a ``single
nucleus approximation'' pioneered in Ref.~\cite{Lattimer91} where
matter was assumed to consist only of neutrons, protons, alpha
particles, and representative heavy nuclei. Multicomponent
calculations of an EOS for supernova simulations have been performed
in several works~\cite{Hix03,Botvina05,
  Sumiyoshi08,Arcones08,Souza09,Hempel10} but these works have not
addressed the matter in the accreted neutron star crust.

\subsection{Quasi-Statistical Equilibrium}

In lieu of a full reaction network, nuclear reactions proceed in
``chunks''; a small chunk of the nuclei present in the current
distribution are assumed to instantaneously undergo a particular
reaction at constant pressure and constant entropy. In the case that
the Helmholtz free energy (enthalpy) per particle is lowered, then the
reaction proceeds and the composition is modified accordingly. At the
temperatures of interest, finite-temperature effects are negligible so
the enthalpy is replaced with the Gibbs energy. Minimizing the Gibbs
energy per particle at fixed pressure rather than the free energy
density has the additional benefit that prevents the system from using
nuclear reactions to unphysically lower the pressure as the baryon
density is increased. As the chunk size approaches zero, this
procedure is equivalent to choosing a quasi-equilibrium state at each
density. The chunk size is chosen to be 1/100th of the total number
density of all nuclei, and this size is sufficiently small to ensure
that the results approximate the correct quasi-equilibrium. This
process does not represent true statistical equilibrium because the
pycnonuclear fusion reactions are not reversible, i.e. fission is not
allowed.

In order to match an old and new configuration at constant pressure,
the nuclear densities and the global quasi-free neutron density,
$n_{n,\mathrm{out}}(1-\phi)$ in the new configuration are scaled by
the same factor (denoted $\alpha$ below) until the pressure matches
that of the original configuration. Neutron emissions and neutron
captures require an additional constraint to ensure that the neutrons
are counted correctly. Denoting $\delta n$ as the decrease in the
number density of parent nuclei due to neutron emission, the number
density of parent nuclei after neutron emission is $\alpha (n_P -
\delta n)$ and the number density of daughter nuclei after neutron
emission is $\alpha (n_D + \delta n)$. Denoting the new configuration
with apostrophes, The number density of
quasi-free neutrons is given by
\begin{equation}
n^{\prime}_{n,\mathrm{out}} (1-\phi^{\prime}) = 
\alpha \left[n_{n,\mathrm{out}} (1-\phi) + \delta n\right] \, ,
\end{equation}
where $\alpha$ is the volume scaling factor and $\phi^{\prime}$ is
implicitly a function of $n^{\prime}_{n,\mathrm{out}}$. The variable
$n^{\prime}_{n,\mathrm{out}}$ must be varied to ensure that this
equation holds. In combination with the requirement that the pressures
are equal, this gives two equations which must be solved for the
variables $\alpha$ and $n^{\prime}_{n,\mathrm{out}}$ to compute the
proper new configuration after any proposed nuclear reaction. This
procedure applies the neutron emission before the compression, but is
equivalent to the opposite choice, i.e. $n_P^{\prime} = \alpha n_P -
\delta n$, in the limit that $\delta n$ is small.

The simulation of a crust begins with an initial composition at a
density near $10^6$ g/cm$^{3}$ and proceeds in a series of
quasi-equilibrium configurations to higher densities. Electron
captures, $\beta$-decays, neutron emissions, and neutron captures are
always allowed if they lower the Gibbs energy per particle. In
practice, since matter is becoming more neutron-rich as the density
increases, electron captures and neutron emission dominate over
$\beta$-decays and neutron captures. Nuclei with number densities less
than $10^{10}$ times the total number density of nuclei are
automatically pruned from the distribution. The heating rate is
determined by computing the change in Gibbs free energy per baryon as
the system proceeds from one configuration to another. In order to
estimate the reduction in heating from neutrino emission in electron
captures, heating from electron captures is multiplied by 1/4 as in
Ref.~\cite{Haensel08}. Because the liquid droplet model is not very
accurate for very light nuclei, all reactions which result in nuclei
with $Z<4$ are not permitted, but this is a good approximation
throughout the crust, as shown below. Nuclei whose neutron or proton
radii are larger than the size of the WS cell are unphysical and thus
also disallowed.

\subsection{Pycnonuclear Fusion Reaction Rates}

Pycnonuclear fusion reaction rates are the most uncertain rates
involved in the accreted crust, in particular the effect of large
neutron skins on rates is not clear~\cite{Horowitz08b}.
Previously, detailed fusion rates have not been widely
available, for example, in \cite{Haensel90}, a simplified fusion rates
were based on approximate S factors obtained from the parameterization
in \cite{Fowler64}. This situation has been improved two-fold: (i)
fusion S-factors for light neutron-rich nuclei were computed in
Ref.~\cite{Beard10}, and (ii) a detailed formalism for fusion rates in a
multicomponent plasma has been described in Ref.~\cite{Yakovlev06}.

Ref.~\cite{Beard10} computes S-factors for $Z=6$, 8, 10, and 12
isotopes with even neutron numbers. In the work below, it is assumed
that Z=4 nuclei always fuse, though this assumption does not
significantly affect our calculations. Nuclei Z=14 are assumed to fuse
whenever Z=12 nuclei fuse, and this assumption does not affect our
results. In order to compute S-factors involving odd proton or neutron
numbers, the proton number is always increased by one and the neutron
number is decreased by one, which slightly decreases the fusion rates.
Ref.~\cite{Beard10} does not compute S-factors for nuclei which are
sufficiently neutron-rich for our study, so it is assumed that the
S-factors for all $Z=6$ nuclei with $A\geq A_{\mathrm{max}}(Z)$ are
the same as that for $A=A_{\mathrm{max}}(Z)$, where
$A_{\mathrm{max}}(Z) \equiv 24$, 28, 40, and 46, for $Z=6$, 8, 10, and
12, respectively. This choice slightly increases the potential for
fusion.

To compute the rates, the formalism in Ref.~\cite{Yakovlev06} is used,
which assumes that the multicomponent plasma is uniformly mixed.
To compute the reaction rates, the parameter $\lambda$ is defined with
\begin{eqnarray}
\lambda_{ij} &=& \frac{A_i+A_j}{A_i A_j Z_i Z_j
\left(Z_i^{1/3}+Z_j^{1/3}\right)} \nonumber \\
&& \times \left[\frac{\rho \left(1-X_n\right) \left<Z\right>}
{\left<A\right> 1.3574 \times 10^{11} 
\mathrm{g}~\mathrm{cm}^{-3}} \right]^{1/3}
\end{eqnarray}
and the plasma temperature
\begin{equation}
T_p^{ij} = \left(\frac{4 \pi Z_i Z_j e^2 n_{ij}}{2 \mu_{ij}}\right)^{1/2}
\end{equation}
where 
\begin{equation}
n_{ij} \equiv \frac{6}{\pi \left[\left(4 \pi n_i/3\right)^{-1/3}+
\left(4 \pi n_j/3\right)^{-1/3}\right]^3}
\end{equation}
for nuclei $i$ and $j$ with number densities $n_i$ and $n_j$, 
and the reduced mass $\mu_{ij} \equiv m_u A_i A_j/(A_i+A_j)$. 
With these definitions the rate is given by 
\begin{eqnarray}
R_{\mathrm{pyc}} &=& 10^{46} C_{\mathrm{pyc}} \frac{8 \rho (1-X_n) x_i
  x_j A_i A_j \left<A\right> Z_i^2 Z_j^2}{\left(1 + \delta_{ij}
\right) (A_i+A_j)^2} \nonumber \\
\times &&
S(E_{pk}) \lambda_{ij}^{3-C_{\mathrm{pl}}}
\exp \left( -\frac{C_{\mathrm{exp}}}{\sqrt{\lambda_{ij}}} \right)
~\mathrm{cm}^{-3}\mathrm{s}^{-1}
\end{eqnarray}
In order to test if a particular fusion reaction is allowed, the the
fusion timescale $n_i/R_{\mathrm{pyc}}$ is compared with the
accretion timescale $y/{\dot{M}}$. If the accretion rate timescale is
larger, than the pycnonuclear reaction is allowed if the fusion will
lower the free energy, otherwise, the fusion is prohibited.

\section{Results}

\begin{figure}
\includegraphics[width=3.4in]{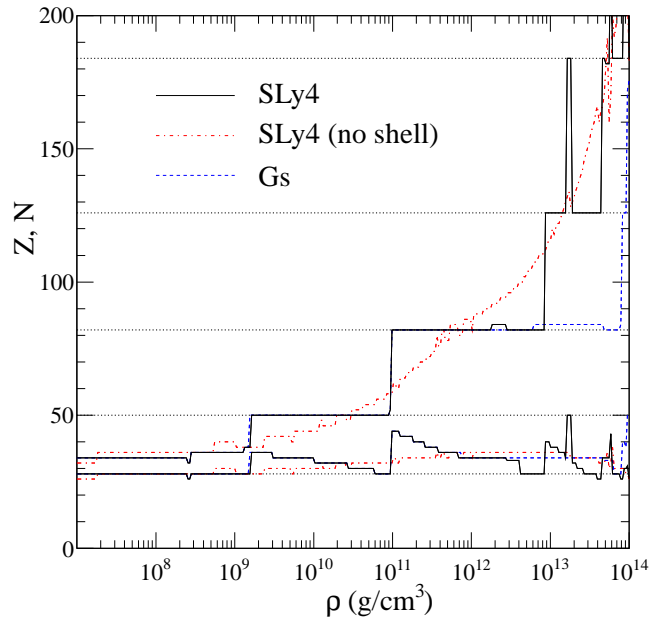}
\caption{(Color online) The composition of the equilibrium neutron
  star crust as a function of density. Deviations originate primarily
  because of the symmetry energy: models where the symmetry energy
  depends more steeply with density have larger Z and smaller N, i.e.
  a composition closer to the valley of stability. The smoother curves
  labeled ``no shell'' do not include shell effects.}
\label{fig:comp}
\end{figure}

It is useful to compare the accreted crust to the much simpler case of
an isolated neutron star crust in full equilibrium (all possible
nuclear reactions are allowed) at zero temperature. The composition in
the equilibrium crust is given in Fig.~\ref{fig:comp}. Except for the
shell effects, the qualitative features agree with recent work in
Ref.~\cite{Newton11}. The neutron magic numbers are indicated by
dashed lines, and it is clear that shell effects dominate the choice
of the neutron number over all densities, while the proton number
steadily decreases to select the proper equilibrium electron fraction
as it decreases with density. The proton number is most often even
because of pairing. Note that at these small temperatures, the single
nucleus approximation is very good because the energy difference
between nuclear masses is typically much larger than the temperature
and because the system is in equilibrium.

\begin{figure}
\includegraphics[width=3.4in]{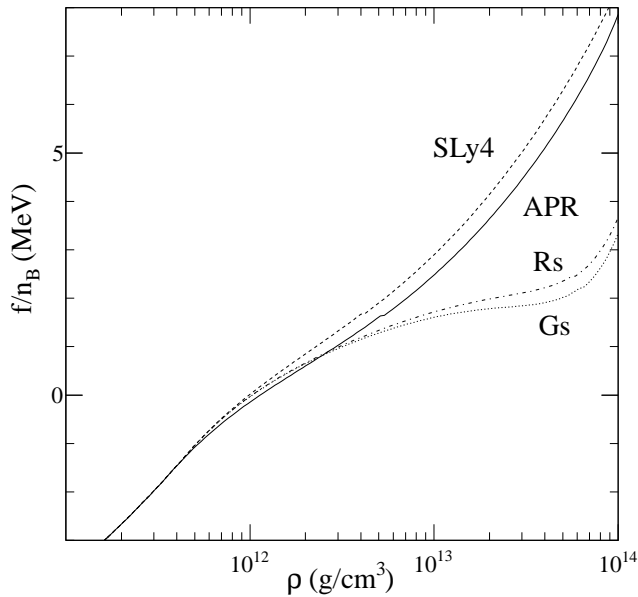}
\caption{The energy per baryon of matter in the cold neutron star
crust in full equilibrium. The distinction between the APR and SLy4
models and the Gs and Rs models is primarily due to the difference
in the density dependence of the nuclear symmetry energy.}
\label{fig:free}
\end{figure}

The free energy per baryon of matter in the full equilibrium is given
in Figure~\ref{fig:free}. Note that the uncertainty in the homogenous
matter EOS affects the free energy per baryon at higher densities even
though the composition is not strongly affected.

\begin{figure*}
\includegraphics[width=3.4in]{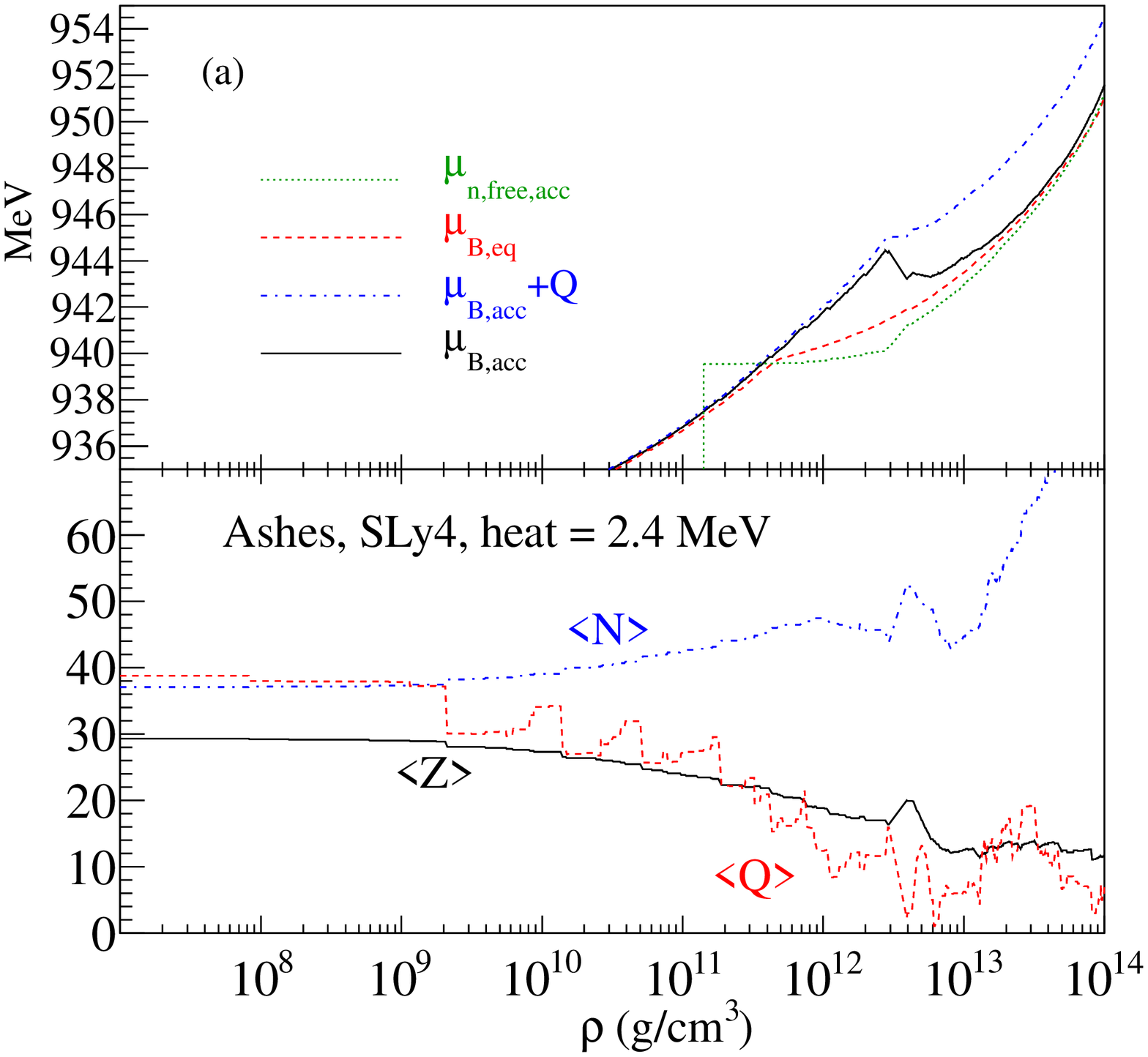}
\includegraphics[width=3.4in]{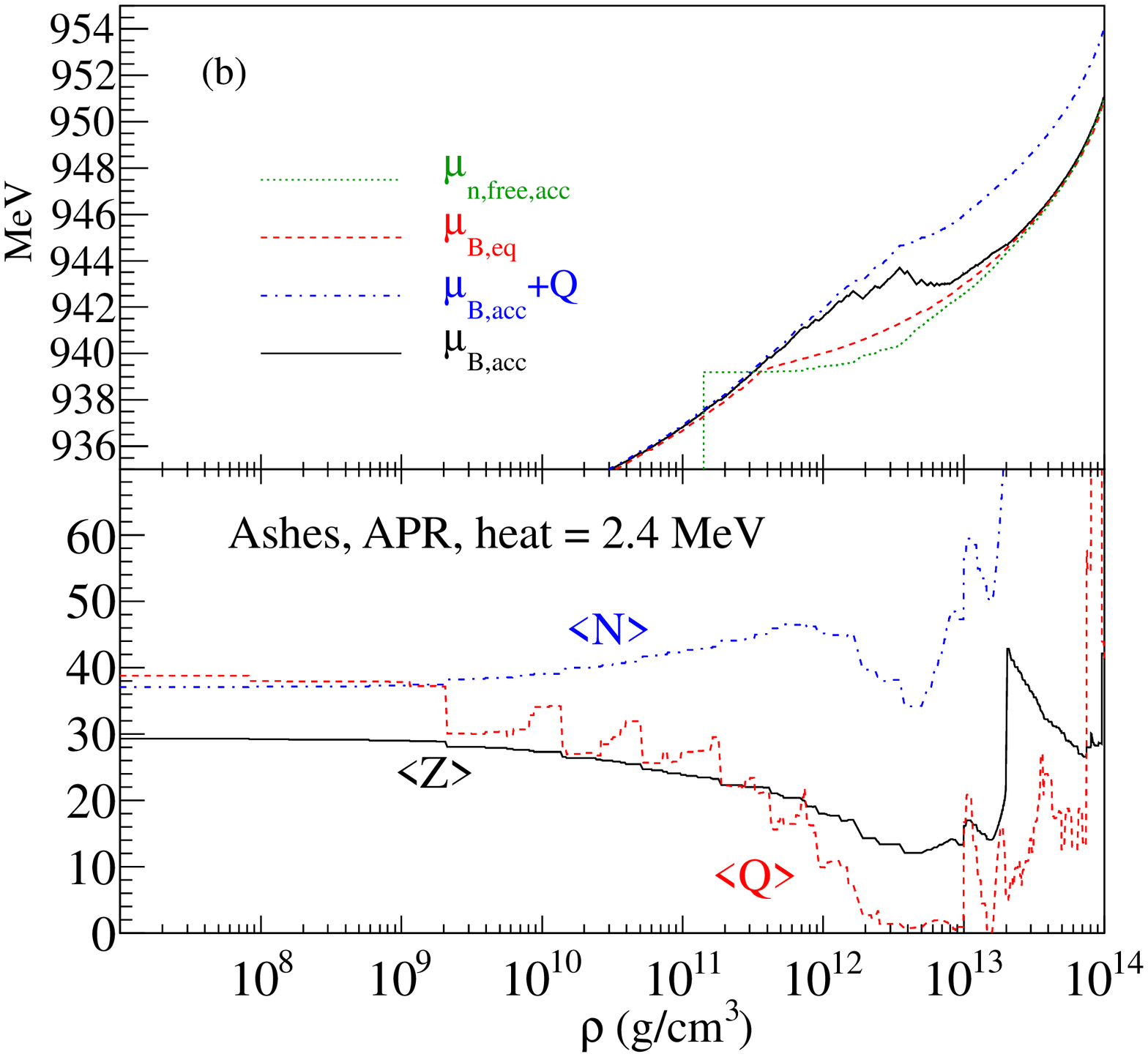}
\includegraphics[width=3.4in]{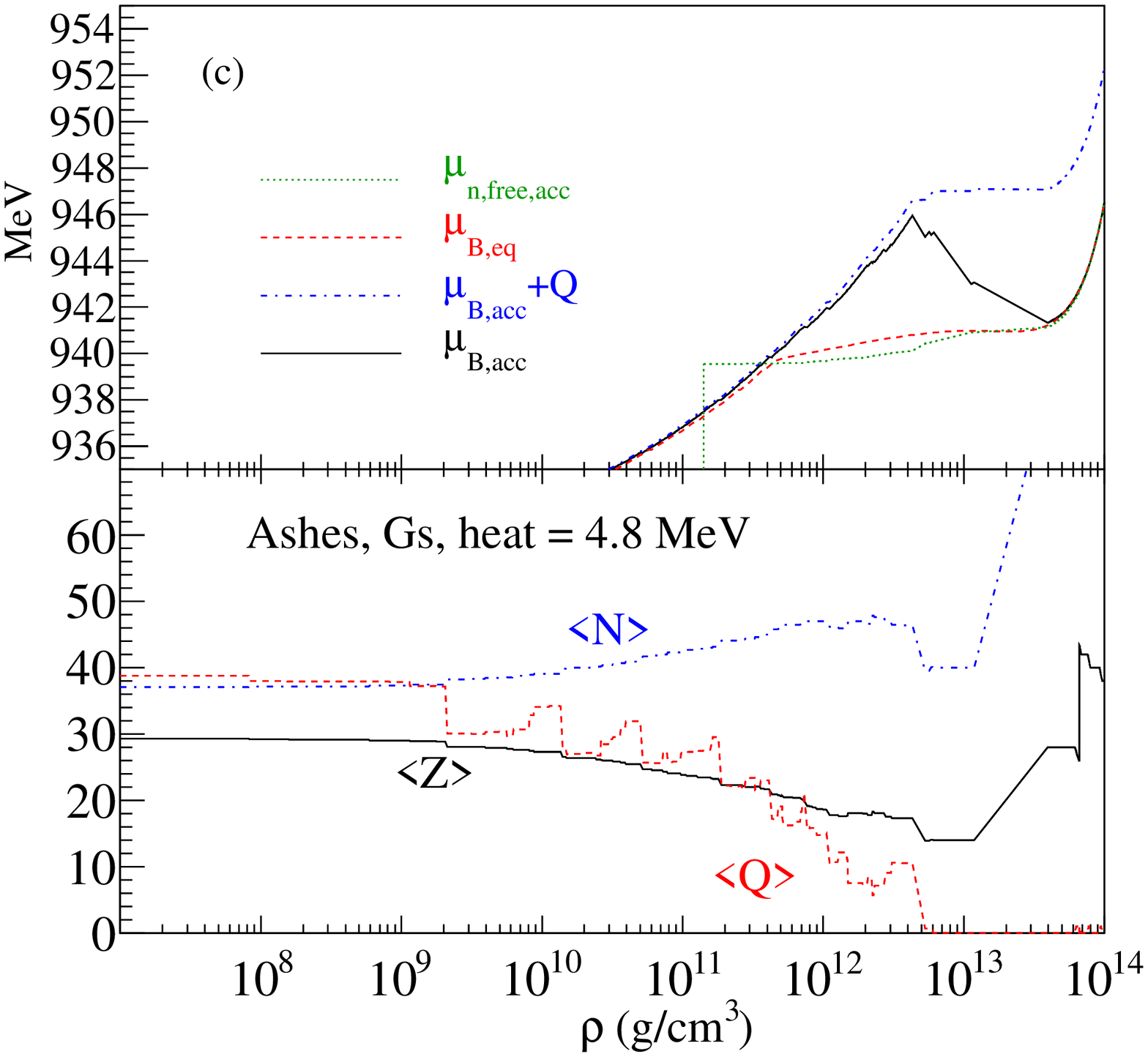}
\includegraphics[width=3.4in]{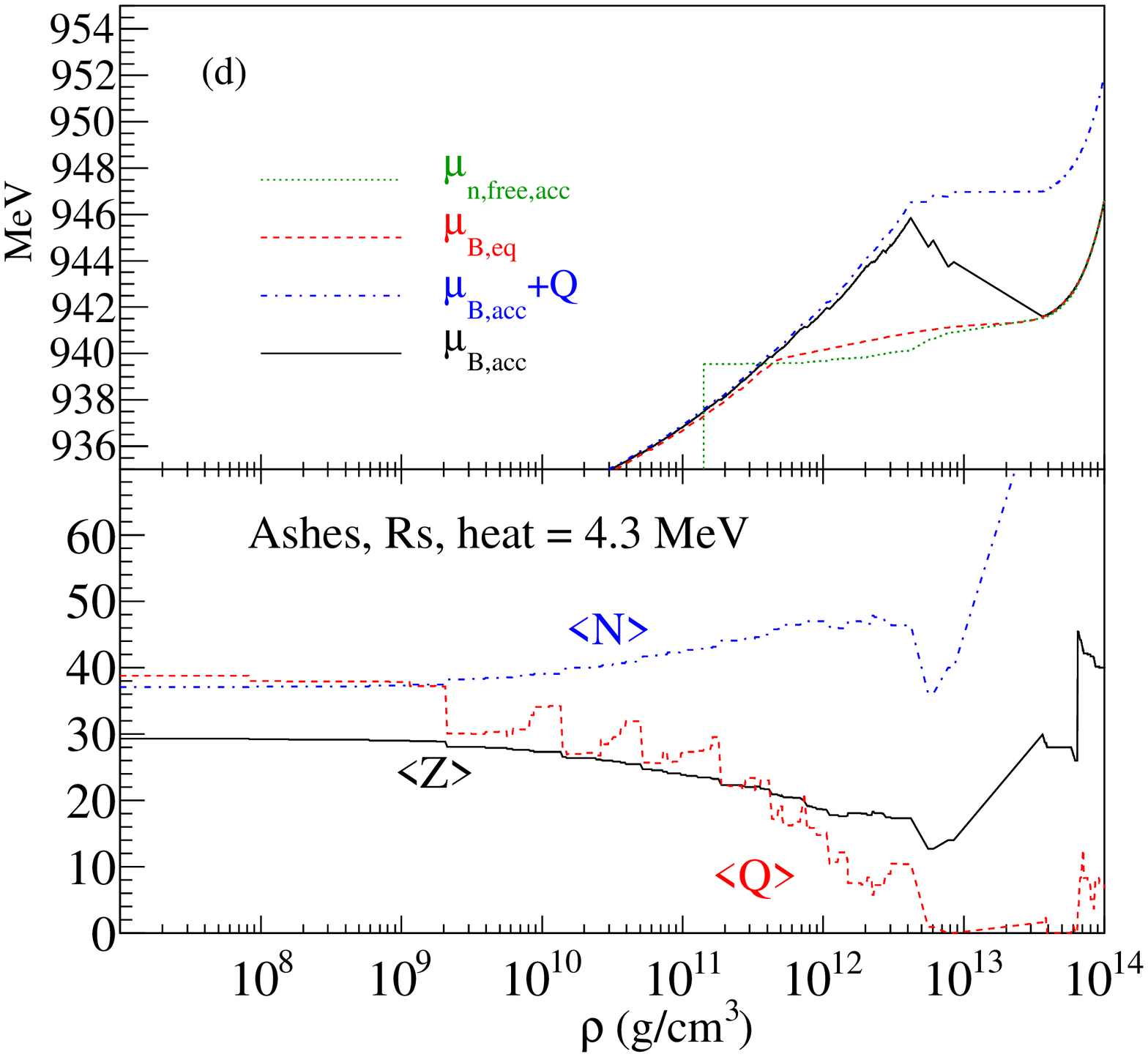}
\caption{(Color online) The accreted crust properties for the four
  models used in this work: (a) SLy4, (b) APR, (c) Gs, and (d) Rs.
  X-ray burst ashes were used as the initial composition in each case.
  Each panel gives the total integrated heat generated by nuclear
  reactions throughout the crust. }
\label{fig:eos}
\end{figure*}

Fig.~\ref{fig:eos} summarizes the properties of the accreted crust for
the four EOSs used in this work. The entire crust is fixed at a
temperature of T=$10^8$ K, and thermal effects do not strongly affect
the composition. The initial composition is taken from Table I in
Ref.~\cite{Horowitz07}. The lower panel for each model shows the
average neutron and proton numbers in the crust as a function of
density along with an impurity parameter, $\left<Q\right>$.
Given the average proton number,
\begin{equation}
\left<Z\right> = \left( \sum_i n_i Z_i \right) \left( \sum_i n_i
\right)^{-1} \, ,
\end{equation}
the impurity parameter is defined as
\begin{equation}
\left<Q\right> = \left[ \sum_i n_i \left( Z_i - \left<Z\right>
\right)^2 \right] \left[ \sum_i n_i \right]^{-1} \, .
\end{equation}
Generally, neutron numbers increase with density and proton numbers
decrease with density, but a large amount of variation is present in
the deepest regions in the crust. The dotted line in the upper panel
is the chemical potential of the quasi-free neutrons in homogeneous
matter at the same number density. The dashed line is the baryon
chemical potential of the full equilibrium crust. The solid line is
the full baryon chemical potential of matter in the accreted crust.
The total amount of heating in the crust is equal to the
baryon chemical potential (the Gibbs free energy per baryon).
because a significant amount of heating occurs at densities slightly 
larger than the neutron drip density, the baryon chemical potential
tends to form a valley at that density where heating overcomes the
natural increase in baryon chemical potential from increasing
pressure. 

The dashed-dotted line is the sum of this baryon chemical potential
with the total integrated heat from all the layers above the current
one. This quantity from Fig. 5 of Ref.~\cite{Haensel08} was defined to
help explain the deep crustal heating phenomenon. The total amount of
deep crustal heating is limited to the energy difference between this
value and the baryon chemical potential in the full equilibrium crust.
In models SLy4 and APR, the amount of deep crustal heat is limited by
the rise of the baryon chemical potential in the full equilibrium
crust. In models Gs and Rs, where the symmetry energy depends less
strongly with density, the baryon chemical potential of the full
equilibrium crust is smaller and thus the deep crustal heating is more
significant.

\begin{figure*}
\includegraphics[width=3.43in]{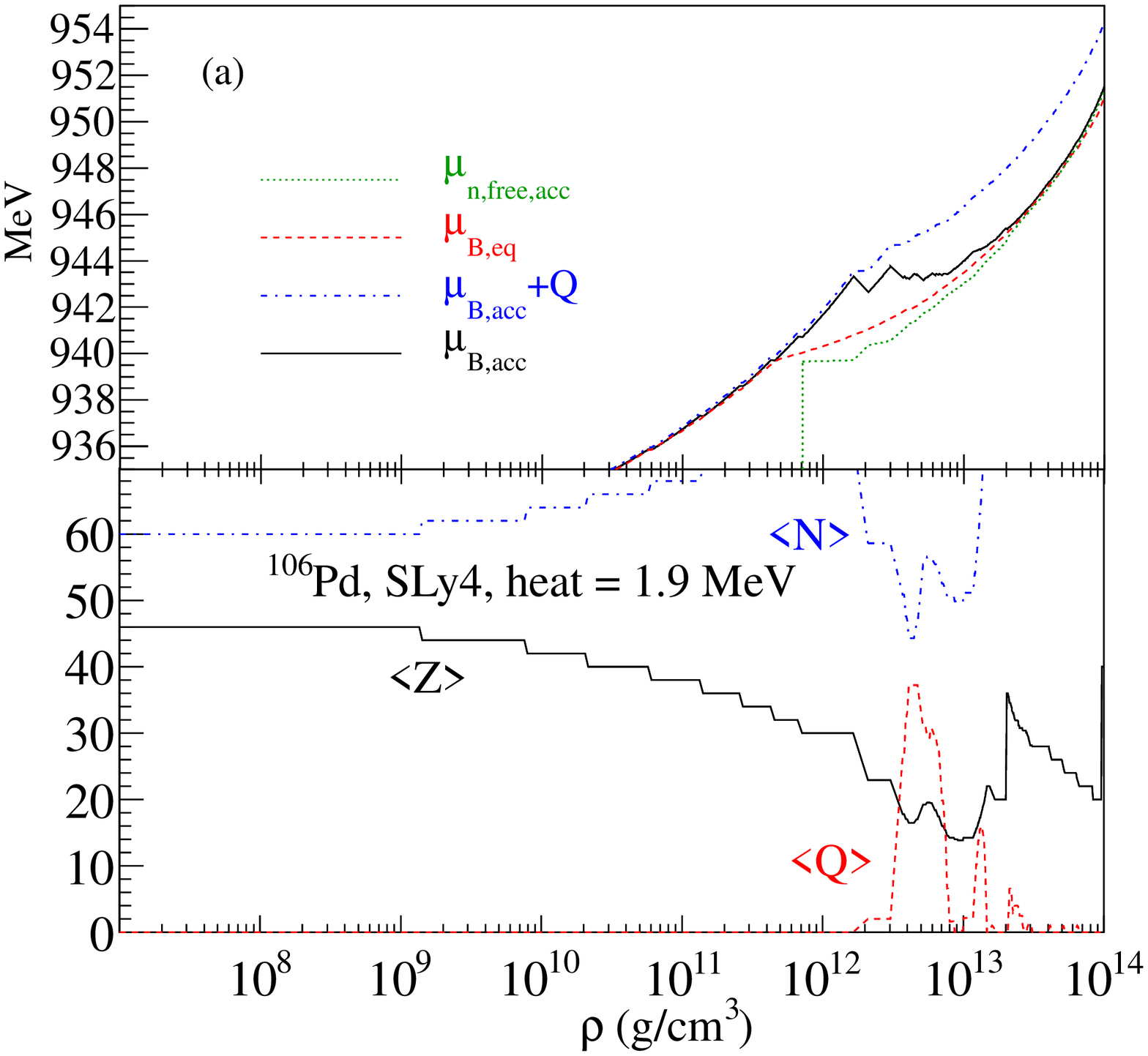}
\includegraphics[width=3.43in]{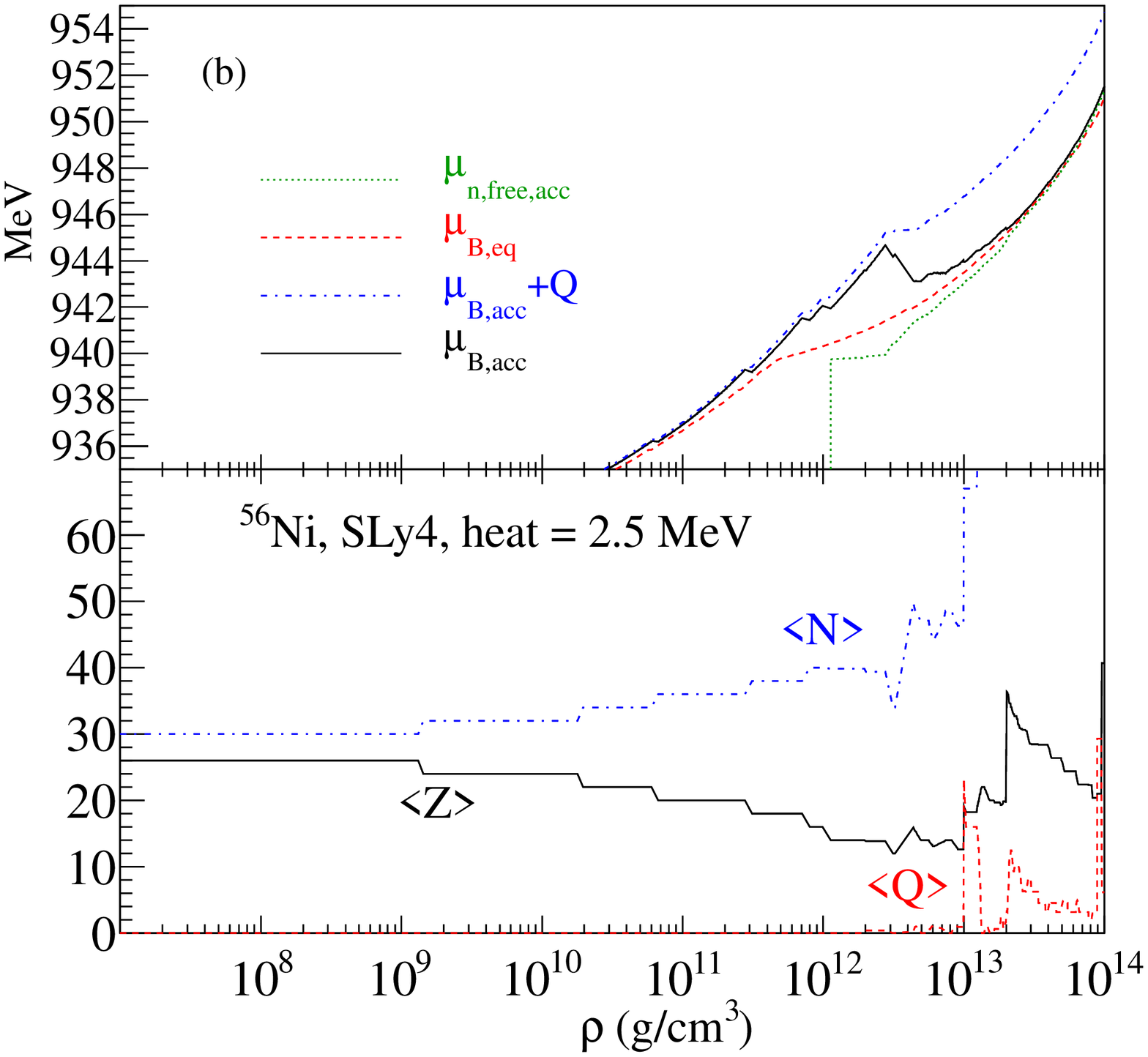}
\caption{(Color online) The accreted crust properties for model SLy4
  but using an initial composition of either pure $^{106}$Pd (a) or
  pure $^{56}$Ni (b). Each panel gives the total integrated heat
  generated by nuclear reactions throughout the crust. }
\label{fig:mods}
\end{figure*}

Variations in the initial composition of matter are explored in the
top part of Fig.~\ref{fig:mods}. The left panel has an initial
composition of pure $^{106}$Pd and the right panel has an initial
composition of pure $^{56}$Ni. The final compositions in these models
are not strikingly different from the X-ray burst ashes used in
Fig.~\ref{fig:eos}. The impurity parameters are much lower, except for
a peak near $4 \times 10^{12}$ g/cm$^{3}$ in the $^{106}$Pd case where
the composition is recovering from the larger number of neutrons
stored in nuclei in that case. Increasing the temperature to $10^9$ K,
or varying the mass accretion rate (important for computing the
relevant accretion timescale to compare with the fusion timescale) by
an order of magnitude did not significantly change the results.

The majority of the heating occurs through the emission of neutrons
near neutron drip. This heating often occurs in large chains which
effectively convert some nuclei entirely into neutrons, leaving the
relative composition of the remaining nuclei unaltered. A sample chain
of this form begins with two $^{40}$Mg nuclei, both of which undergo 6
electron captures and enough neutron emissions to form $^{22}$C. These
two nuclei fuse to form one $^{44}$Mg nucleus, which then may emit four
neutrons to return to the original $^{40}$Mg. This cycle effectively
converts one nucleus entirely into neutrons. Only a fraction of nuclei
can undergo such a cycle at any depth, and thus such cycles are
difficult to resolve in single-nucleus models. While shell closures in
this region are partially softened by the presence of the quasi-free
neutron gas, resolving the nature of shell effects for neutron-rich
nuclei in this region may be helpful in understanding the reaction
pathway details at densities just higher than neutron drip.

\section{Discussion}

This work is an important tool to calibrate more sophisticated network
calculations which follow the evolution of nuclei in the crust as they
begin at the surface and evolve to lower depths. If the
physical nucleon-nucleon interaction leads to a symmetry energy which
depends more weakly with density, then this may result in more heating
and will help alleviate the current difficulties that superburst
models have in reproducing superburst data. Some of this additional
heat, however, will be transmitted to the core rather than the crust,
and a full simulation of the thermal profile which includes 
possible variation in the nuclear masses due to the symmetry energy
is not yet available. 

Compositional nonuniformity may also drive novel heating processes in
the crust~\cite{Medin10a,Medin10b}. While these heating processes are
not included in the current model, they are also fundamentally limited
by the difference in the baryon chemical potential between the
accreted and equilibrium crusts.

The model outlined in this work ignores proton and light-fragment
emission in accreting neutron stars, and may be incorrect near the
crust-core phase transition. Near this density, protons may tunnel
between nuclei and this will also affect the composition. Finally, the
presence of pasta structures and nuclear structure uncertainties in
the masses of nuclei are important in the deepest layers of the crust.
The amount of heat generated at this depth is small, and the
composition at the deepest regions in an accreting neutron star may be
difficult to observe. The composition of the cold crust in equilibrium
might be observable in the giant flares emitted by
magnetars~\cite{Steiner09}.

A one-zone model may not properly estimate the properties of matter
just near neutron drip. To demonstrate this, consider Fig. 5 of
Ref.~\cite{Haensel08}. In the single-nucleus approximation, the baryon
chemical potential, $\mu_B$, drops discontinuously as a function of
the pressure at locations where the composition of the crust changes.
This is clearly unphysical, as this means that the system can lower
its energy by a finite amount by taking a neutron from pressures
slightly lower than the drop in $\mu_B$ to pressures slightly larger
than the drop. This means that neutrons might sink as they are
emitted, and the crust will reconfigure itself to ensure that there
are no strong discontinuities in $\mu_B$. (The electron chemical
potential may also exhibit discontinuities in regions where multiple
electron captures occur.) The multicomponent models in this work soften
these discontinuities considerably, but a flow of neutrons cannot be
ruled out in this one-zone model. A multi-zone generalization of this
work is in progress. 

\section{Acknowledgments:} The author thanks E.F. Brown, A. Cumming,
R. Lau, S. Reddy, and H. Schatz for several useful discussions. This
work was supported by DOE grant DE-FG02-00ER41132, NASA ATFP grant
NNX08AG76G, and by the Joint Institute for Nuclear Astrophysics at MSU
under NSF PHY grant 08-22648.

\bibliographystyle{apsrev}
\bibliography{paper}

\section{Appendix: Gibbs Free Energy Density}

To compute the Gibbs free energy density, it is useful to compute the
chemical potentials and entropies analytically from
Eq.~\ref{eq:totfr}. An advantage of the liquid droplet model formalism
is that these expressions are simple to compute accurately. 

The global quasi-free neutron density is defined with ${\hat n}_n \equiv
n_{n,\mathrm{out}} (1-\phi)$, and this (not $n_{n,\mathrm{out}}$ is
the quantity directly connected to baryon number conservation. Thus,
to compute the Gibbs energy the derivatives
\begin{equation}
{\hat \nu}_n \equiv \left(\frac{\partial f}{\partial {\hat n}_n}
\right)_{\{n_i\},T} 
\end{equation}
and 
\begin{equation}
\nu_i \equiv \left(\frac{\partial f}{\partial n_i}
\right)_{ {\hat n}_n ,\{n_j \forall j\neq i\},T} \,
\end{equation}
are required. The symbol $\nu$ is used rather than $\mu$ which is
reserved for the chemical potentials in infinite matter
\begin{eqnarray}
& \mu_n \equiv \left[ \frac{f_{\infty}(n_{n,\mathrm{out}},T)}
{\partial n_{n,\mathrm{out}}}\right]_{T} \, , \quad \mathrm{and} 
& \nonumber \\
& \mu_e \equiv \left[ \frac{f_{\mathrm{elec}}(n_e,T)}
{\partial n_e}\right]_{T} \, . & 
\end{eqnarray}
Note that these chemical potentials are defined including the rest
mass energy. To be more concise, the subscripts will sometimes be
suppressed in the following. The Gibbs free energy density is then
\begin{equation}
g = {\hat \nu}_n {\hat n}_n + \sum_i \nu_i n_i \, .
\end{equation}
It is useful to note that
$n_{n,\mathrm{in},i}=n_{n,\mathrm{in},i}(\chi_i)$ is distinct quantity
for each nuclear species $i$ and is a function of $\chi_i$. For
simplicity, this is denoted as $n_{ni}$ below. Also note that no
separate term in the Gibbs free energy is required for the electrons
since their number density is not an independent variable. In this
section, it is also easier to think of the function ${\cal
E}_{\mathrm{nuc}}$ as a set of functions of the form ${\cal
E}_{\mathrm{nuc},i}(n_{n,\mathrm{out}},\chi_i)$. For each nucleus $i$,
\begin{equation}
N_i n_e = N_i \left( \sum_j n_j Z_j \right) = Z_i \chi_i n_{ni}
\label{eq:densrel}
\end{equation}
and this implies that $\chi$ and $n_e$ both depend only on the number
densities of nuclei and not separately on $n_{n,\mathrm{out}}$. From
Eq.~\ref{eq:densrel}, one can obtain the relation $ \partial n_e /
\partial n_i = Z_i $ and the derivative
\begin{equation}
\left( \frac{\partial \chi_j} {\partial n_i} 
\right)_{\{n_j\} \forall j \neq i,~ {\hat n}_n} 
=  \frac{N_j Z_i }{Z_j n_{nj} + Z_j 
\chi_j g^{\prime}(\chi_j)} \, .
\end{equation}
This derivative is
non-zero for $i \neq j$ because the nuclei are all coupled by the
Coulomb interaction through the electron density. Some simplification
is afforded by the relation $\phi_i = N_i n_i/ n_{ni}$, where $n_{ni}$
depends only on $\chi_i$, which shows that $\phi_i$ is constant when
all of the $ \{n_i\} $ are held fixed. Thus,
\begin{equation}
\left( \frac{\partial n_{n,\mathrm{out}}} {\partial {\hat n}_n} 
\right)_{\{n_i\}} = (1-\phi)^{-1}
\end{equation}
and
\begin{equation}
\left( \frac{\partial n_{n,\mathrm{out}}} {\partial n_i} 
\right)_{\{n_j\} \forall j \neq i,~{\hat n}_n} =
\frac{{\hat n}_n}{\left(1-\phi\right)^2}
\sum_j \left( \frac{\partial \phi_j}{\partial n_i} \right)
\end{equation}
One can rewrite the derivative
of $\phi_j$ in terms of the derivative of $\chi_j$ determined
above
\begin{equation}
\left( \frac{\partial \phi_j}{\partial n_i} \right)_{\{n_j\} \forall j 
\neq i,~{\hat n}_n} =
\frac{N_j}{n_{nj}} \left[ \delta_{ij} - 
\frac{n_j g^{\prime}(\chi_j)}{n_{nj}} 
\left( \frac{\partial \chi_j}{\partial n_i} \right)_{{\hat n}_n}
\right] \, 
\end{equation} 
where $\delta_{ij}$ is 1 if $i=j$ and zero otherwise. The last
derivative which is required is
\begin{eqnarray}
\left(\frac{\partial N_j^{\prime}}{\partial n_i}
\right)_{\{n_j\} \forall j \neq i,~{\hat n}_n} &=& 
-4 \pi R_{\mathrm{n},j}^2 \left[ \frac{R_{\mathrm{n},j}}{3}
\left( \frac{\partial n_{n,\mathrm{out}}} {\partial n_i} 
\right) \right. \nonumber \\
&& \left. + n_{n,\mathrm{out}} 
\left( \frac{\partial R_{\mathrm{n},j}} {\partial \chi_j} 
\right) \left( \frac{\partial \chi_j} {\partial n_i} 
\right) \right] \, .
\end{eqnarray}
The derivative of $f$ with respect to ${\hat n}_n$ is
\begin{eqnarray}
{\hat \nu}_n &=& \left( \frac{\partial n_{n,\mathrm{out}}} 
{\partial {\hat n}_n} \right) \left\{ \mu_n + \sum_i n_i
\left[ \left(\frac{\partial {\cal E}_{\mathrm{nuc},i}}
{\partial n_{n,\mathrm{out}}}
\right)_{\chi_i} + \right. \right.
\nonumber \\
&& \left. \left. - \frac{4\pi}{3} R_{n,i}^3 m_n 
 \right] \right\} \, .
\end{eqnarray}
Note that this implies
\begin{equation}
{\hat n}_n~\nu_{{\hat n}_n} = n_{n,\mathrm{out}} 
\left(\frac{\partial f}{\partial n_{n,\mathrm{out}}}
\right)_{\{n_i\}} 
 \, .
\end{equation}
The derivative of $f$ with respect to the number density of nucleus $i$ is
\begin{eqnarray}
\nu_i &=& \left(\frac{\partial f}{\partial n_{n,\mathrm{out}}}
\right) \left( \frac{\partial n_{n,\mathrm{out}}} {\partial n_i} 
\right) + Z_i \mu_e  + {\cal E}_{\mathrm{nuc},i} + Z_i m_p \nonumber \\
&& + N_{i}^{\prime} m_n + \sum_j n_j h_{ij}
\end{eqnarray}
where $h_{ij}$ is defined by
\begin{eqnarray}
h_{ij} &\equiv& 
\left(\frac{\partial {\cal E}_{\mathrm{nuc},j}}{\partial \chi_j}
\right)_{n_{n,\mathrm{out}}}
\left( \frac{\partial \chi_j} {\partial n_i} 
\right)  \nonumber \\ &&  + 
\left(\frac{\partial {\cal E}_{\mathrm{nuc},j}}{\partial n_{n,\mathrm{out}}}
\right)_{\chi_j} 
\left( \frac{\partial n_{n,\mathrm{out}}} {\partial n_i} 
\right) + m_n
 \left(\frac{\partial N_j^{\prime}}{\partial n_j}
\right)_{n_{n,\mathrm{out}}}
\end{eqnarray}

\end{document}